\documentstyle[aps,prc]{revtex}
\tighten

\begin{document}
%\draft
%\preprint{HEP/123-qed}

\title{
Induced Proton Polarization for $\pi^0$ Electroproduction\\ 
at $Q^2 = 0.126$ GeV$^2$/c$^2$ around the $\Delta$(1232) Resonance\\
}

%
% Author list for FPP and OOPS collaboration for FPP and N-Delta '95
%   version 1.13                Nov. 30 1997
%

\author{
G.A.~Warren,$^{7,}$\cite{current_warren}
R.~Alarcon,$^1$
C.~Armstrong,$^4$
B.~Asavapibhop,$^{16}$
D.H.~Barkhuff,$^{17,}$\cite{current_barkhuff}
W.~Bertozzi,$^7$ 
V.~Burkert,$^{11}$
J.~Chen,$^7$
J.-P.~Chen,$^{11}$ 
J.R.~Comfort,$^1$ 
D.~Dale,$^{14}$ 
G.~Dodson,$^7$
S.~Dolfini,$^{1,}$\cite{current_dolfini}
K.~Dow,$^7$
M.~Epstein,$^3$
M.~Farkhondeh,$^7$ 
J.M.~Finn,$^4$ 
S.~Gilad,$^7$
R.W.~Gothe,$^2$ 
X.~Jiang,$^{16}$
M.~Jones,$^4$
K.~Joo,$^{7,}$\cite{current_joo}
A.~Karabarbounis,$^{12}$
J.~Kelly,$^{15}$
S.~Kowalski,$^7$ 
C.~Kunz,$^{2,}$\cite{current_barkhuff}
D.~Liu,$^3$ 
R.W.~Lourie,$^{17,}$\cite{current_lourie}
R.~Madey,$^6$
D.~Margaziotis,$^3$ 
P.~Markowitz,$^{15}$
J.I.~McIntyre,$^{4,}$\cite{current_mcintyre}
C.~Mertz,$^1$
B.D.~Milbrath,$^{17,}$\cite{current_milbrath}
R.~Miskimen,$^{16}$ 
J.~Mitchell,$^{11}$
S.~Mukhopadhyay,$^3$
C.N.~Papanicolas,$^{12}$
C.~Perdrisat,$^4$ 
V.~Punjabi,$^8$
L.~Qin,$^9$ 
P.~Rutt,$^{10}$
A.~Sarty,$^5$
J.~Shaw,$^{16}$
S.-B.~Soong,$^7$
D.~Tieger,$^7$
C.~Tschal\ae r,$^7$
W.~Turchinetz,$^7$ 
P.~Ulmer,$^9$
S.~Van~Verst,$^{7,}$\cite{current_vanverst}
C.~Vellidis,$^{12}$ 
L.B.~Weinstein,$^9$
S.~Williamson,$^{13}$
R.J.~Woo,$^{4,}$\cite{current_woo}
A.~Young$^1$\\
(The M.I.T.-Bates OOPS and FPP Collaborations)
}

\address{
 {$^1$  \it Department of Physics and Astronomy, 
                Arizona State University, 
                Tempe, Arizona 85287} \\
 {$^2$  \it Department of Physics, 
                Bonn University, 
                Bonn, Germany} \\
 {$^3$  \it Department of Physics and Astronomy, 
                California State University, 
                Los Angeles, California 90032} \\
 {$^4$  \it Physics Department, 
                College of William and Mary, 
                Williamsburg, Virginia 23187} \\
 {$^5$  \it Department of Physics,
                Florida State University,
		Tallahasse, Florida 32306}\\
 {$^6$  \it Department of Physics, 
                Kent State University, 
                Kent, Ohio 44242} \\
 {$^7$  \it Department of Physics, 
                Laboratory for Nuclear Science and Bates Accelerator Center, 
                Massachusetts Institute of Technology, 
                Cambridge, Massachusetts 02139} \\
 {$^8$  \it Department of Physics and Engineering, 
                Norfolk State University, 
                Norfolk, Virginia 23504} \\
 {$^9$  \it Department of Physics, 
                Old Dominion University, 
                Norfolk, Virginia 23529} \\
 {$^{10}$  \it Department of Physics, 
                Rutgers University, 
                Piscataway, New Jersey 08855} \\
 {$^{11}$ \it Thomas Jefferson National Accelerator Facility, 
                Newport News, Virginia 23606} \\
 {$^{12}$ \it Institute of Accelerating Systems and Applications and 
                Department of Physics, 
                University of Athens, Athens, Greece} \\
 {$^{13}$ \it Physics Department, 
                University of Illinois, 
                Urbana-Champagne, Illinois 61801} \\
 {$^{14}$ \it Department of Physics and Astronomy, 
                University of Kentucky, 
                Lexington, Kentucky 40506}\\ 
 {$^{15}$ \it Department of Physics and Astronomy, 
                University of Maryland, 
                College Park, Maryland 20742} \\
 {$^{16}$ \it Department of Physics, 
                University of Massachusetts, 
                Amherst, Massachusetts 01003} \\
 {$^{17}$ \it Institute for Nuclear and Particle Physics
                and Department of Physics, 
                University of Virginia, \\
                Charlottesville, Virginia 22901}  \\
}

\date{\today}
\maketitle
\begin{abstract}
  We present the first measurement of the induced proton polarization
  $P_n$ in $\pi^0$ electroproduction on the proton around the
  $\Delta$ resonance.  The measurement was made at a central invariant
  mass and a squared four-momentum transfer of $W=1231$ MeV and $Q^2 =
  0.126$ GeV$^2$/c$^2$, respectively.  We measured a large induced
  polarization, $P_n = -0.397 \pm 0.055 \pm 0.009$.  The data suggest
  that the scalar background is larger than expected from a recent
  effective Hamiltonian model.

\end{abstract}


\twocolumn
\narrowtext

At low $Q^2$, the $N\rightarrow\Delta$ transition is dominated by the
magnetic dipole amplitude.  In a simple SU(6) model in which all the
quarks occupy S states in the $N$ and $\Delta$ wavefunctions, the
$N\rightarrow\Delta$ transition is a spin flip of a single quark.  If
the quarks are allowed to occupy D states as well as S states in the
$N$ or $\Delta$ wavefunctions, then electric and Coulomb quadrupole
transitions are allowed~\cite{isgur-karl,drechsel-tiator}. The ratios
of these quadrupole amplitudes to the dominant magnetic dipole
amplitude, referred to as the $R_{EM}$ and $R_{CM}$, are indicative of
the relative importance of the D state in the nucleon and $\Delta$
wavefunction in this model.  

A sensitive probe of the $N\rightarrow\Delta$ transition is pion
production on the free nucleon.  However, many processes in addition
to the $N\rightarrow\Delta$ transition contribute to pion production:
non-resonant nucleon excitation, photon--vector-meson coupling and
excitation of other resonances.  Rescattering of the final-state
hadrons also affects the pion production
observables~\cite{sato-and-lee}.  We refer to the non-resonant
processes as ``background''~\cite{background}.  In order to extract
information about the $N\rightarrow\Delta$ transition from pion
production observables, one must understand the contributions from the
background processes.

Electroproduction experiments were performed in the late 60's and
early 70's in which the $R_{CM}$ was extracted by performing multipole
analysis of ($e$,$e^\prime p$) data acquired over a wide range of
energies and angles~\cite{cmr-measure}. These analyses extracted an
average $R_{CM}$ of roughly $-7\%$ for $Q^2$ up to 1~GeV$^2$/c$^2$.
In 1993, an ($e$,$e^\prime\pi^0$) experiment was conducted at ELSA at
$Q^2 = 0.127$ GeV$^2$/c$^2$~\cite{kalleicher}. The analysis of this
experiment yielded a large $R_{CM}$ of $-0.127 \pm 0.015$, in
agreement with the analysis by Crawford~\cite{crawford} of earlier
($e$,$e^\prime p$) data at the same $Q^2$.  

We conducted a series of $H$($e$,$e^\prime p$)$\pi^0$ measurements at
the same $Q^2$ as the ELSA measurement.  We measured two types of
observables: 1) the cross section over a range of proton scattering
angles with respect to the momentum transfer for a wide range of the
invariant mass around the $\Delta$, 2) the induced proton polarization
in parallel kinematics in which the proton is detected along the
direction of the momentum transfer.  The cross section measurements
allow for the extraction of the $R_{CM}$.  The induced polarization
measurement is sensitive to the background contributions.  We discuss
in this paper the results of the polarization measurement, which is
the first such measurement of the $N\rightarrow\Delta$ transition.

Past electroproduction measurements were performed over a wide range
of $Q^2$, but only the angular dependence of the coincidence cross
section was extracted from the data~\cite{cmr-measure}.  This data
constrains only the real part of the interference response
tensor~\cite{kalleicher}. In parallel kinematics the induced
polarization $P_n$ is proportional to the imaginary part of a
longitudinal-transverse interference response tensor; hence, it is
proportional to the interference of the resonant and background
amplitudes.  In this manner, $P_n$ is sensitive to the same physics
as the beam helicity asymmetry proportional to $R_{LT^\prime}$, the
``fifth response function''~\cite{amaldi}. Thus $P_n$ is in a new class
of pion production observables.

The experiment was conducted in 1995 in the South Hall of
M.I.T.-Bates.  A 0.85\% duty factor, 719~MeV electron beam was
incident on a cryogenic liquid hydrogen target.  Electrons were
detected with the Medium Energy Pion Spectrometer (MEPS)~\cite{meps}
which was located at 44.17$^{\rm o}$ and set at a central momentum of
309~MeV/c.  Coincident protons were detected with the One-Hundred-Inch
Proton Spectrometer (OHIPS)~\cite{ohips} which was located at
-23.69$^{\rm o}$ and set at a central momentum of 674~MeV/c.  The
final-state proton polarization components were measured with the
Focal Plane Polarimeter (FPP)~\cite{barkhuff}.  The central invariant
mass and the squared four-momentum transfer were $W=1231$ MeV and
$Q^2=0.126$ GeV$^2$/c$^2$.  We sampled data over a range of $W$
between 1200 and 1270~MeV.

The focal plane asymmetries were calculated following the procedure
detailed in Ref.~\cite{barkhuff}.  This procedure involved the
use of polarimetry data of elastic scattering from
hydrogen~\cite{milbrath} to determine the false asymmetries of the
polarimeter. In the one photon exchange approximation with unpolarized
electrons, elastically scattered protons cannot be polarized.
Therefore, any measured non-zero polarization is due to false
asymmetries.  The resulting false asymmetries were small, $< 0.004$.

The polarization of the protons at the polarimeter is the asymmetry of
the secondary scattering divided by the $p$-$^{12}C$ inclusive analyzing
power.  We determined the analyzing power by using calibration data of
the FPP taken at the Indiana University Cyclotron
Facility~\cite{lourie93}. From our data taken with an incident proton
energy of 200 MeV and the world's data for analyzing power for
energies between 150 and 300 MeV~\cite{aprile-giboni,mcnaughton}, we
determined a new fit to the functional form of the analyzing power
according to Aprile-Giboni {\em et al.}~\cite{aprile-giboni}.  The
uncertainty in the analyzing power for this measurement was 1.5\%.

In a magnetic spectrometer such as OHIPS, the polarizations at the
target and focal plane are related by a spin precession
transformation. This transformation depends on the precession of the
spin in the spectrometer and on the population of events across the
acceptance.  For this measurement, the transformation simplified to a
simple multiplicative factor for the induced polarization because the
electron beam was unpolarized and the protons were detected along the
direction of the momentum transfer.

To determine this transformation, we used the Monte Carlo program
MCEEP~\cite{mceep} modified to use the spin transfer matrices of
COSY~\cite{cosy}.  We populated events across the acceptance using a
preliminary electroproduction model by Sato and Lee (SL) based on their
photoproduction model described in Ref.~\cite{sato-and-lee}.  The
transformation was
\begin{equation}
P_n = (-1.070 \pm 0.016) P_X,
\end{equation}
where $P_X$ is the polarization component extracted from the azimuthal
asymmetry of the secondary scattering, and $P_n$ is the normal type
polarization at the target.  We varied parameters in the COSY and
MCEEP models by their measured uncertainties to determine the
uncertainty of the spin precession transformation.  

To compare to theoretical models, we corrected the measured
polarization for finite acceptance effects.  We determined the
correction with MCEEP using the SL pion production model:
\begin{equation}
\frac{P_n {\rm\ for\ Point\ Acceptance}}
     {P_n {\rm\ for\ Full\ Acceptance}} = 1.159 \pm 0.011.
\end{equation}
This correction is mostly due to the large electron acceptance.  The
uncertainty in the acceptance correction reflects uncertainties in the
experimental acceptance.

Applying the spin transformation factor and the acceptance correction,
we determined that the induced polarization for a point acceptance was
\begin{equation}
P_n = -0.397 \pm 0.055 \pm 0.009,
\end{equation}
where the first uncertainty is statistical and the second is
systematic.  Our analysis does not depend on the absolute scale of the
model predictions.  Thus, the smooth variations of the cross section
and of the induced polarization over the experimental phase space
predicted by the model of Sato and Lee suggest that the model
sensitivity should be sufficiently small to be neglected for this
measurement.  Corrections to $P_n$ due to radiative processes are
small, 0.02\%, and were not included.

In parallel kinematics all the response functions can be constructed
from two complex amplitudes which we label $S$ and
$T$~\cite{raskin-donnelly}.  In terms of the Chew-Goldberger-Low-Nambu
amplitudes~\cite{cgln} and multipole amplitudes expanded up to p
wave~\cite{amaldi}, these two amplitudes are
\begin{equation}
\begin{tabular}{c c c c c c c c c c c c}
\label{eqn:s_and_t}
$S$ & $=$ & $F_5^\prime - F_6^\prime$ & $\approx$ 
    & $S_{0+}$ & $-$ & $S_{1-}$ & $-$ & $4S_{1+}$, &     &\\
$T$ & $=$ & $F_1 + F_2$ & $\approx$ 
    & $E_{0+}$ & $+$ & $M_{1-}$ & $-$ & $3E_{1+}$ & $-$ & $M_{1+}$.
\end{tabular}
\end{equation}

In parallel kinematics, $P_n$ is proportional to the imaginary part of a
longitudinal-transverse interference divided by the unpolarized cross
section~\cite{raskin-donnelly}.  In terms of $S$ and $T$,
\begin{eqnarray}
\label{eqn:pn_fact}
P_n & = & \frac{-\sqrt{2\epsilon_s(1+\epsilon)}{\rm Im}S^*T}{|T|^2 
                  + \epsilon_s|S|^2}, \\
    & = & \frac{-\sqrt{2\epsilon_s(1+\epsilon)}
                  \left( \beta_S  - \zeta_S\beta_T\right)}
               {\left(1 + \beta_T^2\right) 
                + \epsilon_s\left( \beta_S^2 + \zeta_S^2\right)},
\end{eqnarray}
where $\epsilon=(1+2q_{lab}^2/{Q^2}\cdot
tan^2\frac{1}{2}\Theta_e)^{-1}$, $\epsilon_s =
{Q^2}/{q_{cm}^2}\cdot\epsilon$, $q_{lab}$ ($q_{cm}$) is the
three-momentum transfer in the lab (center-of-momentum) frame,
$\Theta_e$ is the scattering angle of the electron with respect to the
beam, $\beta_{S(T)} = {\rm Re}S(T)/{\rm Im}T$ and $\zeta_S = {\rm Im}
S/{\rm Im} T$.  

The $0^{th}$ order approximation to $P_n$ is obtained by assuming only
a purely resonant $N\rightarrow\Delta$ transition contributes. Then at
resonance, the contributing amplitudes are purely imaginary, and thus
\begin{equation}
\beta_S=\beta_T=0\ \ {\rm and}\ \ \zeta_S = 4\frac{R_{CM}}{1+3\cdot R_{EM}}.
\label{eqn:beta_S}
\end{equation}
This approximation gives $P_n = 0$.  A non-zero $\beta_S$ and/or
$\beta_T$ at resonance, comes from background contributions.
In this manner, $P_n$ is sensitive to the background.

In Fig.~\ref{fig:pn_model} our result is compared to two different
pion production models plotted over a range of the invariant mass $W$
at a fixed $Q^2 = 0.126$ (GeV/c)$^2$. Results from a preliminary
electroproduction model based off the published SL photoproduction
model~\cite{sato-and-lee} are plotted for 0\% and 1.4\% probability of
a D state in the $\Delta$ wavefunction.  The model of Mehrotra and
Wright for the simultaneous fit to $\pi^0$ and $\pi^+$ production data
requiring unitarity (MW)~\cite{mehrotra} is also plotted.  This model
does not consider $\Delta$ resonance quadrupole amplitudes.  Neither
of these models successfully reproduces the measured $P_n$.

The constraints on the ratios due to this measurement are illustrated
in Fig.~\ref{fig:beta_s}.  The two bands denote the regions of
$\{\beta_T,\beta_S\}$ consistent with this measurement for $\zeta_S =
0$ and $-0.4$.  These values of $\zeta_S$ correspond to an $R_{CM}=0$
and $R_{CM} = -9.1\%$ when calculated from Eq.~(\ref{eqn:beta_S}) with
$R_{EM}=-3\%$.  Also shown are the $\{\beta_T,\beta_S\}$ points for
the SL and MW models.  The SL model with a deformed (non-deformed)
$\Delta$ has $\zeta_S = 0.001(-0.047)$, which violates the simple
relation of Eq.~(\ref{eqn:beta_S}) because of a strong imaginary
$S_{0+}$.  The MW model does not consider imaginary scalar
contributions so that $\zeta_S \equiv 0$.  Since $\zeta_S$ of these
models are approximately zero, the points on the graph should be
compared to the vertically-hatched region.

For the wide range of $\beta_T$ and $\zeta_S$ in the figure, $\beta_S$
is larger than 20\%.  It is possible to satisfy the restrictions of
$P_n$ with lower $\beta_S$, but this requires $\zeta_S < -0.4$.
However, the sum $\zeta_S^2 + \beta_S^2$ is limited by the small
longitudinal contribution to the cross section~\cite{baetzner}.  The
results from the companion cross section data will provide additional
information to constrain the ratios.

For the SL model to describe the $P_n$ data, the two extreme
corrections to the model are to increase either $\beta_S$ or
$-\zeta_S\beta_T$.  As the model differs from the measurement by
roughly a factor of two, we want to signficantly change the ratios.
Since the model describes the measured cross section as a function of
the invariant mass well~\cite{cross_section}, we cannot radically
alter the transverse contributions.  $\zeta_S$ differs by only 0.05
between the SL calculations with non-deformed and deformed $\Delta$,
so any large change in the real or imaginary scalar amplitudes must come
from non-resonant contributions.  Following these conjectures, we
conclude that the large $P_n$ of this measurement indicates that the
scalar background contributions are larger than expected from the SL
model.

The inclusion of rescattering in the SL model has a significant effect
on the scalar background contributions compared to the MW model.
Both models use a similar description of the Born amplitudes at the
tree level: pseudovector $\pi NN$ coupling and $\rho$ exchange.
However, the real scalar contributions are quite different as
demonstrated by the difference in the $\beta_S$ values in
Fig.~\ref{fig:beta_s}.  Thus, the rescattering procedure in the SL
model significantly enhances the background scalar contributions.

It is difficult to directly compare the background of this
measurement with that of measurements from which the $R_{CM}$ is
extracted.  In general, the two observables can involve different
combinations of multipole amplitudes.  In addition, $P_n$ is sensitive
to the real part of the background, whereas the observables used to
extract $R_{CM}$ are sensitive to the imaginary part.

Previous extractions of the $R_{CM}$ neglected the non-resonant terms
under the assumption that they are small.  Our data demonstrate that
the background contributions are significant compared to the dominant
resonant contributions and are not well described by recent models.
Therefore, one cannot {\em a priori} neglect the background terms in the
$R_{CM}$ extraction.

In summary, we measured a large induced polarization for pion
production at $W=1231$~MeV and $Q^2 = 0.126$~GeV$^2$/c$^2$.  The data
suggest that the scalar background is larger than expected from recent
effective Hamiltonian models.  We demonstrated that the large induced
polarization of this measurement provides a significant constraint on
scalar background contributions.  Results from the companion
M.I.T.-Bates cross sections measurements and from future experiments
planned at several facilities will constrain theoretical approaches
and improve our understanding of the $N\rightarrow\Delta$
transition.

%%% Acknowledgments

The authors wish to thank the staff of M.I.T.-Bates as well as
T.-S. H. Lee for his preliminary model calculations.  This work was
supported in part by the U.S. Department of Energy and the
U.S. National Science Foundation.

%
% Begin Figures
%

%\newpage

\begin{figure}
%\vspace{2.2in}
%\special{psfile=pn_model.ps hscale=40 vscale=40 hoffset=0 voffset=-310}
%\special{psfile=pn_model.ps hscale=40 vscale=40 hoffset=0 voffset=-70}
\caption{Comparison of measured $P_n$ with the two models.  The
  dot-dash is from MW.  The long (short) dash is from SL with a 0\%
  (1.4\%) probability of a D state in the $\Delta$ wavefunction.  The
  solid line at $P_n = 0$ is the approximation of no background
  contributions.  The uncertainty shown is the statistical and
  systematic uncertainties added in quadrature.}
\label{fig:pn_model}
\end{figure}
%\vspace{1 in}
%\newpage

\begin{figure}
%\vspace{2.0in}
%\special{psfile=beta_s.ps hscale=40 vscale=40 hoffset=0 voffset=-320}
%\special{psfile=beta_s.ps hscale=40 vscale=40 hoffset=0 voffset=-80}
\caption{Regions of $\beta_S$ and $\beta_T$ consistent with this
  measurement for two values of $\zeta_S$.  The region filled with
  vertical (diagonal) lines correspond to $\zeta_S = 0.0 (-0.4)$. The
  regions for each $\zeta_S$ denote the one standard deviation
  uncertainty in the constraint.  The solid circle (square) indicates
  the $\{\beta_T,\beta_S\}$ of the SL model for a deformed
  (non-deformed) $\Delta$.  The empty circle is for the MW
  model.  See text for a further description.}
\label{fig:beta_s}
\end{figure}                                                             
%\vspace{2.0in}

% 
% Begin References
%


\begin{references}

%\vspace{-0.5 in}

\bibitem[*]{current_warren}
Present Address: Dept. f\"ur Physik und
        Astronomie, Universit\"at Basel, CH-4056 Basel, Switzerland.

\bibitem[\dag]{current_barkhuff}
Present address: Laboratory for Nuclear Science, Massachusetts
Institute of Technology, Cambridge, MA 02139.

\bibitem[\ddag]{current_dolfini}
Present address: Hughes Missile Systems, Tucson, AZ 85750. 

\bibitem[\S]{current_joo}
Present address: Thomas Jefferson National Accelerator Facility,
Newport News, VA 23606.

\bibitem[\|]{current_lourie}
Present address: Dept. of Physics and Astronomy, State University of
New York Stony Brook, Stony Brook, NY 11794.

\bibitem[\P]{current_mcintyre}
Present address: Serin Physics Laboratory, Rutgers University,
Piscataway, NJ 08855.

\bibitem[**]{current_milbrath}
Present address: Dept. of Physics and Astronomy, Eastern Kentucky
University, Richmond, KY 40475.

\bibitem[\ddag\ddag]{current_vanverst} 
Present address: Washington State Department of Health, Olympia, WA
98504.

\bibitem[\S\S]{current_woo}
Present address: TRIUMF, Vancouver, British Columbia, Canada V6T 2A3.

\bibitem{isgur-karl}
N. Isgur and G. Karl, Phys. Rev. D {\bf 18}, 4187 (1978).

\bibitem{drechsel-tiator}
D. Drechsel and L. Tiator, J. Phys. {\bf G18}, 449 (1992).

\bibitem{sato-and-lee}
T. Sato and T.-S. H. Lee, Phys. Rev. C {\bf 54}, 2660 (1996).

\bibitem{background}
C. N. Papanicolas, in {\it Topical Workshop on Excited Baryons},
edited by G. Adams, N.C. Mukhopadhyay and P. Stoler 
(World Scientific, Singapore, 1989).

 \bibitem{cmr-measure}
%C. Mistretta {\it et al.}, Phys. Rev. {\bf 184}, 1487 (1969).
%R. L. Crawford, Nucl. Phys. {\bf B28}, 573 (1971).
%J. C. Alder {\it et al.}, Nucl. Phys. {\bf B46}, 573 (1972).
%R. Siddle {\it et al.}, Nucl. Phys. {\bf B35}, 93 (1972).
%K. Batzner {\it et al.}, Nucl. Phys. {\bf B76}, 1 (1974).
F. Foster and G. Hughes, Rept. Prog. Phys. {\bf 46}, 1445 (1983).

\bibitem{kalleicher} 
%F. Kalleicher {\it et al.}, Z. Phys. A {\bf 359}, 201 (1997).
F. Kalleicher, U. Dittmayer, R. W. Gothe, H. Putsch, T. Reichelt, 
B. Schoch and M. Wilhelm, Z. Phys. A {\bf 359}, 201 (1997).

\bibitem{crawford}
R. L. Crawford, Nucl. Phys. {\bf B28}, 573 (1971).

\bibitem{amaldi} 
E. Amaldi, S. Fubini and G. Furlan, {\it Pion Electroproduction:
Springer Tracts in Modern Physics {\bf 83}} (Springer, Berlin, 1979).

\bibitem{meps}
K. I. Blomqvist, MEPS Design Report, Bates Internal Report 78-02 (1978).

\bibitem{ohips}
S. Turley, Massachusetts Institute of Technology, Ph. D. Thesis (1984).

\bibitem{barkhuff}
D. H. Barkhuff, University of Virginia, Ph. D. Thesis (1996).

\bibitem{milbrath}
B. Milbrath {\it et al.}, Phys. Rev. Lett. {\bf 80}, 452 (1998).

\bibitem{lourie93} 
R. W. Lourie {\it et al.}, I.U.C.F. Scientific and Technical Report, 
        {\bf 135} (April 1993).

\bibitem{aprile-giboni}
%A. Waters {\it et al.}, Nucl. Inst. Meth. {\bf 153}, 401 (1978).
%D. Besset {\it et al.}, Nucl. Inst. Meth. {\bf 166}, 379 (1979).
%R. D. Ransome {\it et al.}, Nucl. Inst. Meth. {\bf 201}, 309 (1982).
%E. Aprile-Giboni {\it et al.}, Nucl. Inst. Meth. {\bf 215}, 147 (1983).
E. Aprile-Giboni, R. Hausammann, E. Heer, R. Hess, C. Lechanoine-Le Luc, 
W. Leo, S. Morenzoni, Y. Onel and D. Rapin, 
Nucl. Inst. Meth. {\bf 215}, 147 (1983).

\bibitem{mcnaughton}
M. W. McNaughton {\it et al.}, Nucl. Inst. Meth. {\bf A241}, 435 (1985).

\bibitem{mceep}
P. E. Ulmer, {\it MCEEP - Monte Carlo for Electro-Nuclear
        Coincidence Experiments}, CEBAF-TN-91-01 (1991).

\bibitem{cosy}
M. Berz, Nucl. Inst. Meth. {\bf A298}, 473 (1990);
M. Berz, {\em COSY INFINITY}, Los Alamos LA-11857-C:137, (1990).

\bibitem{raskin-donnelly}
A. S. Raskin and T. W. Donnelly, Ann. Phys. {\bf 191}, 78 (1989).
The definitions of the multipole amplitudes in the above article differ from
those we used, which are defined in Ref.~\cite{amaldi}. 

\bibitem{cgln}
G. F. Chew, M. L. Goldberger, F. E. Low and Y. Nambu, 
Phys. Rev. {\bf 106}, 1345 (1957).

\bibitem{mehrotra}
S. Mehrotra and L. E. Wright, Nucl. Phys. {\bf A362}, 461 (1981).

\bibitem{baetzner} 
K. Baetzner {\it et al.}, Phys. Lett. B {\bf 39}, 575 (1972).
 
\bibitem{cross_section}
C. Mertz and C. Vellidis, to be published.  C. Vellidis, TJNAF 
$N\rightarrow\Delta$ Workshop, 1997.


\end{references}
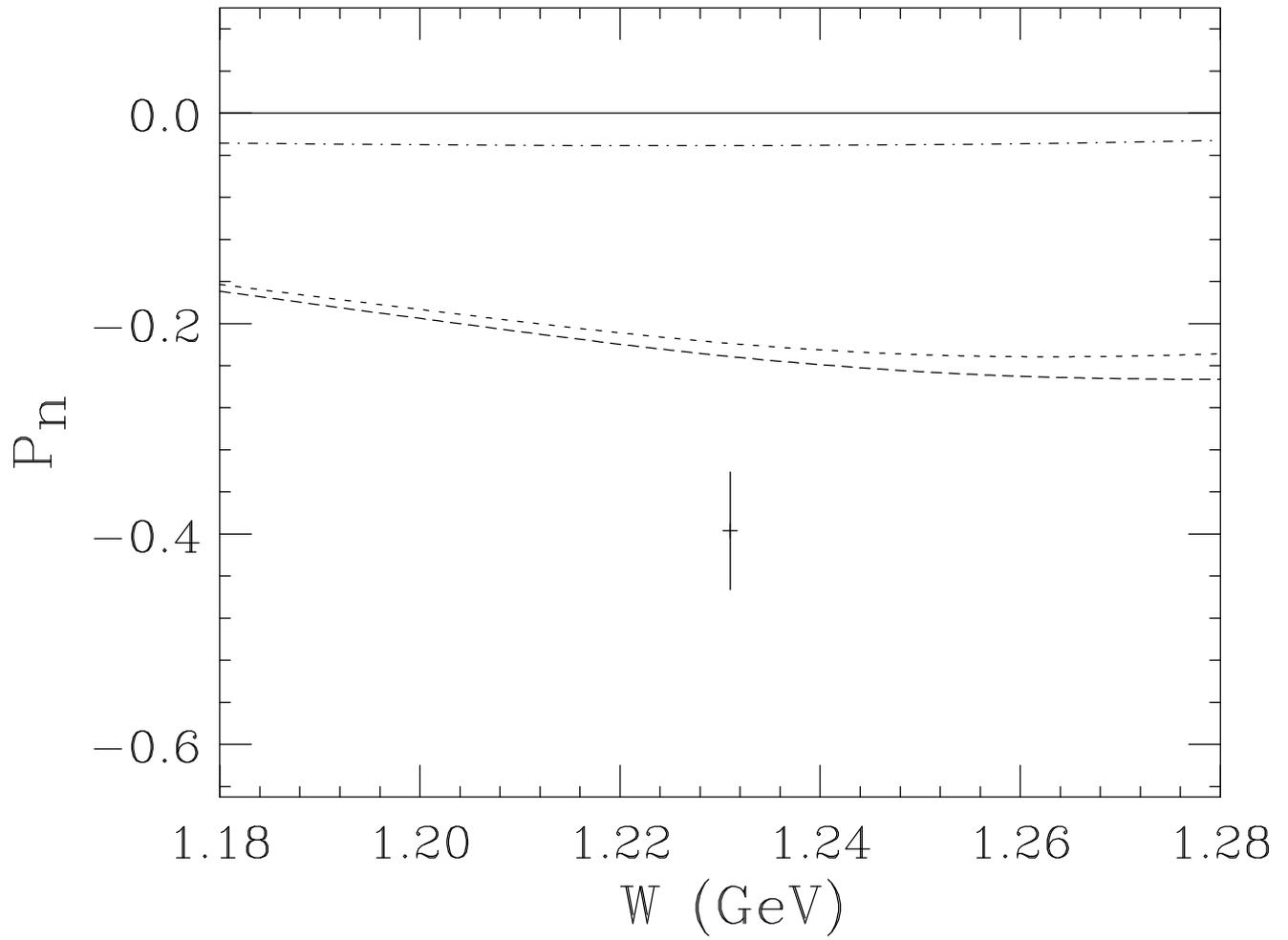
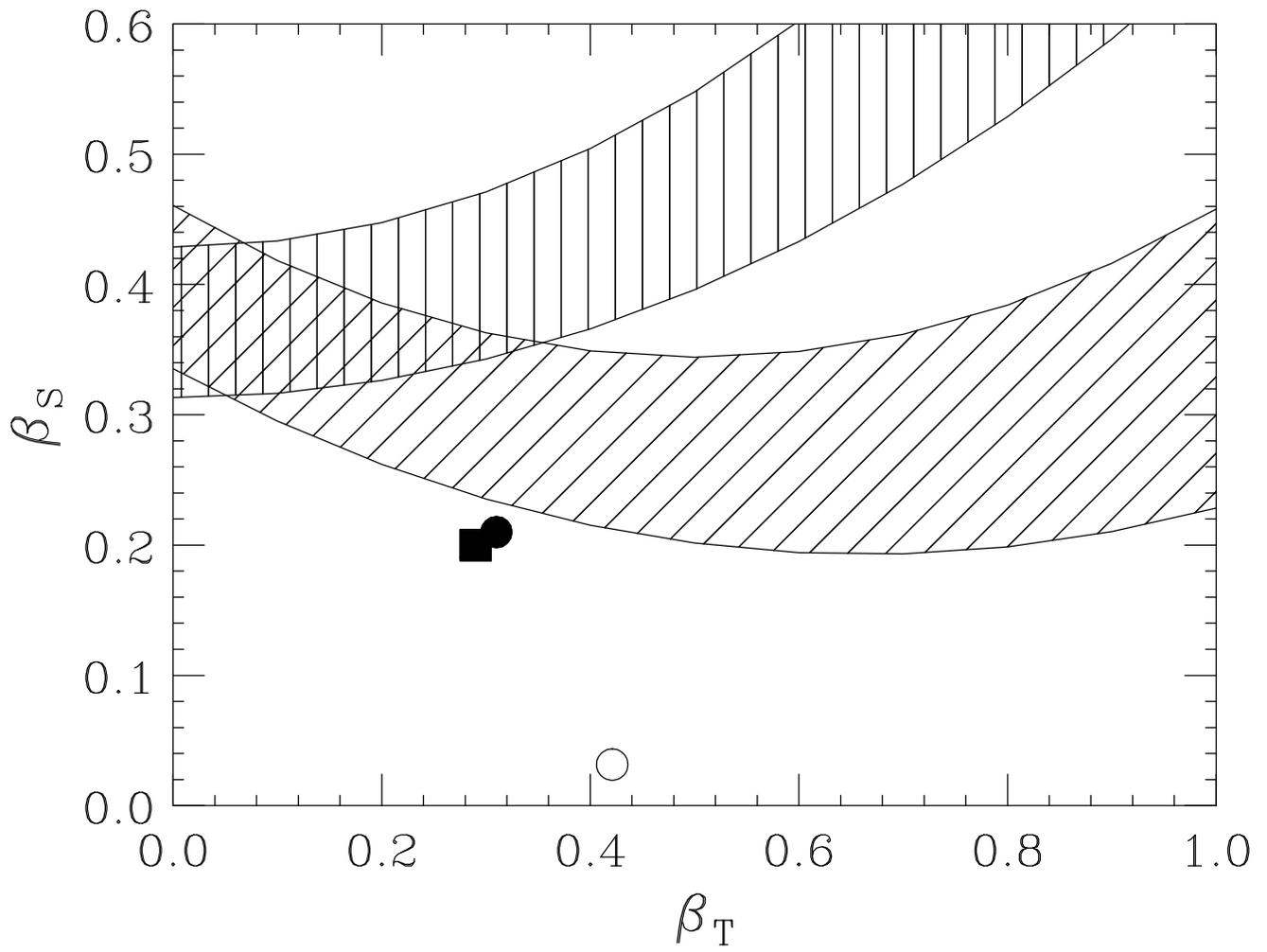
\end{document}